\documentclass[twocolumn,showpacs,amsmath,amssymb,10pt]{revtex4}
\usepackage{amsfonts}
\usepackage{bm}

\newcommand{\ot}{{\,\otimes\,}}
\newcommand{{\Cd}}{{\mathbb{C}^d}}

\def\oper{{\mathchoice{\rm 1\mskip-4mu l}{\rm 1\mskip-4mu l}%
{\rm 1\mskip-4.5mu l}{\rm 1\mskip-5mu l}}}
\def\<{\langle}
\def\>{\rangle}

\newtheorem{DEF}{Definition}

\newtheorem{lemma}{Lemma}

\begin{document}
\title{\textbf{Constructing new optimal entanglement witnesses}} \author{Dariusz
Chru\'sci\'nski, Justyna Pytel and Gniewomir Sarbicki\thanks{email:
darch@phys.uni.torun.pl} }
\affiliation{Institute of Physics, Nicolaus Copernicus University,\\
Grudzi\c{a}dzka 5/7, 87--100 Toru\'n, Poland}

\begin{abstract}

We provide a new class of indecomposable entanglement witnesses. In
$4 \times 4$ case it reproduces  the well know Breuer-Hall witness.
We prove that these new witnesses are optimal and atomic, i.e. they
are able to detect the ``weakest" quantum entanglement encoded into
states with positive partial transposition (PPT).  Equivalently, we
provide a new construction of indecomposable atomic maps in the
algebra of $2k \times 2k$ complex matrices. It is shown that their
structural physical approximations give rise to entanglement
breaking channels. This result supports recent conjecture by Korbicz
{\em et. al.}

\end{abstract}

\maketitle

\section{Introduction}

The interest on  quantum entanglement has dramatically increased
during the last two decades due to the emerging field of quantum
information theory \cite{QIT}. It turns out that quantum
entanglement may be used as basic resources in quantum information
processing and communication. The prominent examples are quantum
cryptography, quantum teleportation, quantum error correction codes
and quantum computation.

Since the quantum entanglement is the basic resource for the new
quantum information technologies it is therefore clear that there is
a considerable interest in efficient theoretical and experimental
methods of entanglement detection (see \cite{HHHH} for the review).

The most general approach to characterize quantum entanglement uses
a notion of an entanglement witness (EW)  \cite{EW1,EW2}. A
Hermitian operator $W$ defined on a tensor product
$\mathcal{H}=\mathcal{H}_A \ot \mathcal{H}_B$ is called  an EW iff
1) $\mbox{Tr}(W\sigma_{\rm sep})\geq 0$ for all separable states
$\sigma_{\rm sep}$, and 2) there exists an entangled state $\rho$
such that $\mbox{Tr}(W\rho)<0$ (one says that $\rho$ is detected by
$W$). It turns out that a state is entangled if and only if it is
detected by some EW \cite{EW1}. There was a considerable effort in
constructing and analyzing the structure of EWs
\cite{Terhal2,O,Lew1,Lew2,Lew3,Bruss,Toth,Bertlmann,Brandao,Gniewko,how}.
In fact, entanglement witnesses have been measured in several
experiments \cite{EX,Wu}. Moreover, several procedures for
optimizing EWs for arbitrary states were proposed \cite{O,O1,O2,O3}.
It should be stressed that there is no universal $W$, i.e. there is
no entanglement witness which detects all entangled states. Each
entangled state $\rho$ may be detected by a specific choice of $W$.
It is clear that each EW provides a new separability test and it may
be interpreted as a new type of Bell inequality \cite{W-Bell}. There
is, however, no general procedure for constructing  EWs.

Due to the Choi-Jamio{\l}kowski isomorphism \cite{Choi,Jam} any EW
corresponds to a linear positive map $\Lambda :
\mathcal{B}(\mathcal{H}_A) \rightarrow  \mathcal{B}(\mathcal{H}_B)$,
where by $\mathcal{B}(\mathcal{H})$ we denote the space of bounded
operators on the Hilbert space $\mathcal{H}$. Recall that a linear
map $\Lambda$ is said to be positive if it sends a positive operator
on $\mathcal{H}_A$ into a positive operator on $\mathcal{H}_B$. It
turns out \cite{EW1} that a state $\rho$ in $\mathcal{H}_A \ot
\mathcal{H}_B$ is separable iff $(\oper_A \ot \Lambda)\rho$ is
positive definite for all positive maps $\Lambda :
\mathcal{B}(\mathcal{H}_B) \rightarrow \mathcal{B}(\mathcal{H}_A)$
(actually this result is based on \cite{Woronowicz1}).
Unfortunately, in spite of the considerable effort, the structure of
positive maps is rather poorly understood
\cite{Woronowicz1,Stormer1,Choi1,Woronowicz2,Robertson,Tang,TT,Osaka,Benatti,
Kye,Ha,Kossak1,Hall,Breuer,OSID-W,atomic,CMP,kule}.

In the present paper we provide a construction of a new class of
positive maps in $\mathcal{B}(\mathbb{C}^{2k})$ with $k\geq 2$. Our
construction uses the well-known reduction map as a building block.
It turns out that for $k=2$ our construction reproduces Breuer-Hall
maps \cite{Breuer,Hall} but for $k>2$ it gives completely new family
of maps. It is shown that proposed maps are indecomposable (i.e.
they are able to detect entangled PPT states) and atomic (i.e. they
are able to detect ``weakly'' entangled PPT states). As a byproduct
we construct new families of PPT entangled states detected by our
maps.

The paper is organized as follows: for pedagogical reason we collect
basic definitions and introduce the most important properties of
positive maps and entanglement witnesses in Section~\ref{DEF}.
Section~\ref{RED} provides basic construction. Then in
Section~\ref{PROP} we study basic properties of our maps/witnesses
(indecomposability, atomicity, optimality). Section~\ref{SPA}
discusses structural physical approximation (SPA)
\cite{SPA1,SPA2,SPA3} of our maps. It is shown that the
corresponding SPA gives rise to entanglement breaking channels and
hence it supports recent conjecture by Korbicz {\em et. al.}
\cite{SPA3}. Final conclusions are collected in the last Section.

\section{Positive maps, entanglement witnesses and all that}
\label{DEF}

For the reader convenience we recall basic definitions and
properties which are important throughout this paper.

\subsection{Positive maps}

Let $\Lambda : \mathcal{B}(\mathcal{H}_A) \rightarrow
\mathcal{B}(\mathcal{H}_B)$ be a positive linear map. In what
follows we shall consider only finite dimensional Hilbert spaces
such that  ${\rm dim}\mathcal{H}_A =d_A$ and ${\rm dim}\mathcal{H}_B
=d_B$. One calls $\Lambda$ $k$-positive if
\begin{equation}\label{}
    \oper_k \ot \Lambda : M_k \ot \mathcal{B}(\mathcal{H}_A) \longrightarrow
M_k \ot \mathcal{B}(\mathcal{H}_B)\ ,
\end{equation}
is positive. In the above formula $M_k$ denotes a linear space of $k
\times k$ complex matrices and $\oper_k : M_k \rightarrow M_k$ is an
identity map, i.e. $\oper_k(A) = A$ for each $A \in M_k$. A positive
map which is $k$-positive for each $k$ is called completely positive
(CP). Actually, if $d_A,d_B < \infty$ one shows \cite{Choi} that
$\Lambda$ is CP iff it is $d$-positive with $d = \min\{d_A,d_B\}$.

\begin{DEF} A positive map $\Lambda$ is decomposable if
\begin{equation}\label{}
    \Lambda = \Lambda_1 + \Lambda_2 \circ T\ ,
\end{equation}
where $\Lambda_1$ and $\Lambda_2$ are CP and $T$ denotes
transposition in a given basis. Maps which are not decomposable are
called indecomposable (or nondecomposable).
\end{DEF}
\begin{DEF} A positive map $\Lambda$ is atomic  if it cannot be
represented as
\begin{equation}\label{}
    \Lambda = \Lambda_1 + \Lambda_2 \circ T\ ,
\end{equation}
where $\Lambda_1$ and $\Lambda_2$ are 2-positive.
\end{DEF}
\begin{DEF} A positive map $\Lambda$ is optimal if and only if for any CP map
$\Phi$, the map $\Lambda - \Phi$ is no longer positive.
\end{DEF}

\subsection{Entanglement witnesses}

Using Choi-Jamio{\l}kowski isomorphism \cite{Choi,Jam} each positive
map $\Lambda$ gives rise to entanglement witness $W$
\begin{equation}\label{CJ}
    W = d_A (\oper_A \ot \Lambda)P^+_A\ ,
\end{equation}
where $P^+_A$ denotes maximally entangled state in $\mathbb{C}^{d_A}
\ot \mathbb{C}^{d_A}$ and $\oper_A$ denotes an identity map acting
on $\mathcal{B}(\mathcal{H}_A)$. One has an obvious
\begin{DEF} An entanglement witness $W$ defined by (\ref{CJ})
is decomposable/indecomposable (atomic) [optimal] $\{$ $k$--EW $\}$
if and only if the corresponding positive map $\Lambda$ is
decomposable/indecomposable (atomic) [optimal] $\{$ $k$--positive
$\}$.
\end{DEF}

It is clear that $W \in \mathcal{B}(\mathcal{H}_A \ot
\mathcal{H}_B)$ is a decomposable EW iff
\begin{equation}\label{Wd}
    W = A + B^\Gamma\ ,
\end{equation}
where $A,B \geq 0$ and $B^\Gamma = (\oper_A \ot T)B$ denotes partial
transposition. Witnesses which cannot  be represented as in
(\ref{Wd}) are indecomposable

Let $\psi$ be a normalize vector in $\mathcal{H}_A \ot
\mathcal{H}_B$. Denote by ${\rm SR}(\psi)$ the number of nonvanishig
Schmidt coefficients of $\psi$. One has
\begin{equation}\label{}
    1 \leq {\rm SR}(\psi) \leq d \ .
\end{equation}
Now, $W$ is $k$--EW iff
\begin{equation}\label{}
    \< \psi |W|\psi\> \geq 0 \ ,
\end{equation}
for each $\psi$ such that ${\rm SR}(\psi) \leq k$. Evidently, $W\geq
0$ iff $W$ is $d$--EW. Now, $W$ is atomic if it cannot be
represented as
\begin{equation}\label{}
    W = W_1 + W_2^\Gamma\ ,
\end{equation}
where $W_1$ and $W_2$ are 2--EWs. Finally, $W$ is optimal EW iff for
any $P\geq 0$, $W-P$ is no longer EW. Following \cite{O} one has the
following criterion for the optimality of $W$: if the set of product
vectors $\psi \ot \phi \in \mathcal{H}_A \ot \mathcal{H}_B$
satisfying
\begin{equation}\label{}
    \< \psi \ot \phi|W|\psi\ot \phi\>=0\ ,
\end{equation}
span the total Hilbert space $\mathcal{H}_A \ot \mathcal{H}_B$, then
$W$ is optimal.

\subsection{Detecting quantum entanglement}

Positive maps and EWs are basic tools in detecting quantum
entanglement. A state $\rho$ in $\mathcal{H}_A \ot \mathcal{H}_B$ is
separable if and only if for all positive maps $\Lambda :
\mathcal{B}(\mathcal{H}_B) \rightarrow \mathcal{B}(\mathcal{H}_A)$
one has
\begin{equation}\label{}
    (\oper_A \ot \Lambda)\rho \geq 0 \ .
\end{equation}
Equivalently, iff for each entanglement witness $W$
\begin{equation}\label{}
    {\rm Tr}(\rho W) \geq 0\ .
\end{equation}
Note that entangled PPT states can be detected by indecomposable
maps/witnesses only. Let  $\sigma$ be a density operator in
$\mathcal{H}_A \ot \mathcal{H}_B$. Following \cite{SN} one
introduces its Schmidt number
\begin{equation}\label{SN-rho}
    \mbox{SN}(\sigma) = \min_{p_k,\psi_k}\, \left\{ \,
    \max_{k}\, \mbox{SR}(\psi_k)\, \right\}  \ ,
\end{equation}
where the minimum is taken over all possible pure states
decompositions
\begin{equation}\label{}
    \sigma = \sum_k \, p_k\, |\psi_k\>\<\psi_k|\ ,
\end{equation}
with $p_k\geq 0$, $\sum_k\, p_k =1$ and $\psi_k$ are normalized
vectors in $\mathcal{H}_A \ot \mathcal{H}_B$. Note, that if $\sigma
= |\psi\>\<\psi|$, then ${\rm SN}(\sigma) = {\rm SR}(\psi)$. Again
one has $ 1 \leq {\rm SN}(\sigma) \leq d$. Suppose now that $\sigma$
is PPT but entangled. Intuitively, the `weakest'' quantum entangled
encoded in $\sigma$ corresponds to the situation when ${\rm
SR}(\sigma) = {\rm SR}(\sigma^\Gamma) = 2$. Such ``weakly''
entangled PPT states can be detected by atomic maps/witnesses only.

\section{Reduction map as a building block}  \label{RED}

Let us start with an elementary positive map in
$\mathcal{B}(\mathbb{C}^n)$ called reduction map \cite{Horodecki}
\begin{equation}\label{Rn}
    R_n(X)=\mathbb{I}_{n}\mathrm{Tr}X-X\ ,
\end{equation}
for $X \in \mathcal{B}(\mathbb{C}^n)$. It is well known that $R_n$
is completely co-positive (i.e. $R_n \circ T$ is CP) and hence
optimal. Recently, this map was generalized by Breuer and Hall
\cite{Breuer,Hall} to the following family of positive maps
\begin{equation}\label{BH}
 \Phi^U_{2k}(X) = \frac{1}{2(k-1)}\Big( R_{2k}(X) - UX^TU^\dagger\Big) \ ,
\end{equation}
where $U$ is an arbitrary antisymmetric unitary $2k \times 2k$
matrix. It was shown  that these map are indecomposable
\cite{Breuer,Hall} and optimal \cite{Breuer}. Such  antisymmetric
unitary matrix  may be easily construct as follows
\begin{equation}\label{}
    U = V U_0 V^\dagger\ ,
\end{equation}
where $V$ stands for real orthogonal matrix $(VV^\dagger =
VV^T=\mathbb{I}_{2k})$ and
\begin{equation}\label{}
    U_0 = \mathbb{I}_k \ot J\ ,
\end{equation}
with $J$ being  $2 \times 2$  symplectic matrix
\begin{equation}\label{J}
 J=\left( \begin{array}{c c} 0 & 1 \\ -1 & 0\end{array}
    \right)\ .
\end{equation}
It is therefore clear that in this case one has
\begin{equation}\label{BH1}
 \Phi^U_{2k}(X) = V \Phi^0_{2k}(V^\dagger X V) V^\dagger \ ,
\end{equation}
where $\Phi^0_{2k}$ corresponds to $\Phi^U_{2k}$ with $U=U_0$.
Actually, one can always find a basis in $\mathbb{C}^{2k}$ such that
$U$ takes the "canonical form" $U_0$. Interestingly for $k=2$ the
Breuer-Hall map $\Phi^0_4$ reproduces well known Robertson map
\cite{Robertson} who provided it as an example of an extremal (and
hence optimal)  indecomposable positive map. Moreover, Robertson
construction may be nicely described in terms of $R_2$ as follows
\cite{atomic}
\begin{widetext}
\begin{equation}\label{R4}
    \Phi^0_4\left( \begin{array}{c|c} X_{11} & X_{12} \\ \hline X_{21} & X_{22} \end{array}
\right) = \frac 12 \left( \begin{array}{c|c} \mathbb{I}_2\,
\mbox{Tr} X_{22} &  -[X_{12} + R_2(X_{21})] \\ \hline  -[X_{21} +
R_2(X_{12})] & \mathbb{I}_2\, \mbox{Tr} X_{11}
\end{array} \right) \ ,
\end{equation}
\end{widetext}
where $X_{kl} \in \mathcal{B}(\mathbb{C}^2)$. This pattern is
reproduced for arbitrary $k$. It is easy to show that the action of
$\Phi^0_{2k}$ may be represented as follows:
\begin{widetext}
\begin{eqnarray} \label{Phi-2k}
\Phi^0_{2k}\left( \begin{array}{c|c|c|c} X_{11} & X_{12} & \cdots &
X_{1k}
 \\ \hline X_{21} & X_{22} & \cdots & X_{2k} \\ \hline \vdots & \vdots & \ddots & \vdots \\
 \hline X_{k1} & X_{k2} & \cdots & X_{kk}\end{array} \right) = \frac{1}{2(k-1)}\left( \begin{array}{c|c|c|c}
\mathbb{I}_{2}(\mathrm{Tr}X-\mathrm{Tr}X_{11}) &
 -(X_{12}+R_2(X_{21})) & \cdots & -(X_{1k}+R_2(X_{k1})) \\ \hline  -(X_{21}+R_2(X_{12})) &
 \mathbb{I}_{2}(\mathrm{Tr}X-\mathrm{Tr}X_{22}) & \cdots & -(X_{2k}+R_2(X_{k2})) \\
 \hline \vdots & \vdots & \ddots & \vdots \\ \hline
 -(X_{k1}+R_2(X_{1k})) & -(X_{k2}+R_2(X_{2k})) & \cdots & \mathbb{I}_{2}(\mathrm{Tr}X-\mathrm{Tr}X_{kk}) \end{array}
 \right)\ ,
\nonumber
\end{eqnarray}
\end{widetext}
where again $X_{ij}$ are $2 \times 2$ blocks. Hence $\Phi^0_{2k}$ is
defined in (\ref{BH}) by $R_{2k}$ but the above pattern shows that
it basically uses reduction map $R_2$ only. We stress that $R_2$ is
exceptional: it is not only optimal but also extremal. Indeed, the
corresponding entanglement witness $W_2 = 2(\oper \ot R_2)P^+_2$
reads as follows
\begin{equation}\label{}
    W_2 = \mathbb{I}_2 \ot \mathbb{I}_2 - P^+_2 =
    \left( \begin{array}{c c|c c} \cdot & \cdot & \cdot & -1 \\ \cdot & 1 & \cdot & \cdot \\
    \hline \cdot & \cdot & 1 & \cdot \\ -1 & \cdot & \cdot & \cdot \end{array}
    \right)\ .
\end{equation}
and $W_2 = P^\Gamma$, where $P=|\psi\>\<\psi|$ with
\begin{equation}\label{}
    |\psi\> = |01\> - |10\>\ .
\end{equation}
Note, that $R_2$ may be nicely represented  as follows
\begin{equation}\label{R-J}
    R_2(X)=J X^T J^\dagger \ ,
\end{equation}
with $J$ defined in (\ref{J}). Formula (\ref{R-J}) provides Kraus
representation for $R_2 \circ T$ and shows that $R_2$ is completely
co-positive.

In the present paper we propose another construction of maps in
$\mathcal{B}(\mathbb{C}^{2k})$. Now, instead of treating a $2k
\times 2k$ matrix $X$ as a $k \times k$ matrix with $2\times 2$
blocks $X_{ij}$  we consider alternative possibility, i.e. we
consider $X$ as a $2 \times 2$ with $k \times k$ blocks and define
\begin{widetext}
\begin{equation}\label{Psi-2k}
    \Psi^0_{2k}\left( \begin{array}{c|c} X_{11} & X_{12} \\ \hline X_{21} & X_{22} \end{array}
\right) = \frac 1k \left( \begin{array}{c|c} \mathbb{I}_k\,
\mbox{Tr} X_{22} &  -[X_{12} + R_k(X_{21})] \\ \hline  -[X_{21} +
R_k(X_{12})] & \mathbb{I}_k\, \mbox{Tr} X_{11}
\end{array} \right) \ .
\end{equation}
\end{widetext}
Again, normalization factor guaranties that the map is unital, i.e.
$\Psi^0_{2k}(\mathbb{I}_2 \ot \mathbb{I}_k) = \mathbb{I}_2 \ot
\mathbb{I}_k$. It is clear that for $k=2$ one has
\begin{equation}\label{}
    \Phi^0_4 = \Psi^0_4 \ .
\end{equation}
We stress that our new construction is much simpler than
$\Phi^0_{2k}$ and it uses as a building block the true reduction map
in $\mathcal{B}(\mathbb{C}^k)$. Moreover, it is clear that it
provides a natural generalization of the original Robertson map in
$\mathcal{B}(\mathbb{C}^4)$.

Now, our task is to prove that $\Psi^0_{2k}$ defines a positive map.
It is enough to show that each rank-1 projector $P$ is mapped via
$\Psi^0_{2k}$ into a positive element in
$\mathcal{B}(\mathbb{C}^{2k})$, that is, $\Psi^0_{2k}(P)\geq 0$. Let
$P = |\psi\>\<\psi|$ with arbitrary $\psi$ from $\mathbb{C}^{2k}$.
Now, due to $\mathbb{C}^{2k} = \mathbb{C}^k \oplus \mathbb{C}^k$ one
has
\begin{equation}\label{}
    \psi = \psi_1 \oplus \psi_2\ ,
\end{equation}
with $\psi_1,\psi_2 \in\mathbb{C}^k$ and hence
\begin{equation}\label{}
    P = \left( \begin{array}{c|c} X_{11} & X_{12} \\ \hline X_{21} & X_{22} \end{array}
\right) = \left( \begin{array}{c|c} |\psi_1\>\<\psi_1| & |\psi_1\>\<\psi_2| \\
\hline |\psi_2\>\<\psi_1| & |\psi_2\>\<\psi_2| \end{array}
     \right)   \  .
\end{equation}
One has therefore
\begin{equation}\label{}
    \Psi^0_{2k}(P) = \frac 1k \left( \begin{array}{c|c} \mathbb{I}_k\, \mbox{Tr}
X_{22} & - A \\ \hline  - A^\dagger & \mathbb{I}_k\, \mbox{Tr}
X_{11} \end{array} \right) \ ,
\end{equation}
where the linear operator $A : \mathbb{C}^k \rightarrow
\mathbb{C}^k$ reads as follows
\begin{equation}\label{}
    A = |\psi_1\>\<\psi_2| - |\psi_2\>\<\psi_1| + \<
    \psi_1|\psi_2\>\, \mathbb{I}_k \ .
\end{equation}
Let $\<\psi_j|\psi_j\> = a_j^2 > 0$ (if one of $a_j$ vanishes then
evidently one has $\Psi^0_{2k}(P)\geq 0$). Defining
\begin{equation}\label{}
    L = \sqrt{k} \left( \begin{array}{c|c} \mathbb{I}_k\, a_2^{-1}
 & \mathbb{O}_k \\ \hline  \mathbb{O}_k & \mathbb{I}_k\, a_1^{-1}
\end{array} \right) \ ,
\end{equation}
one finds
\begin{equation}\label{}
   L \Psi^0_{2k}(P) L^\dagger =  \left( \begin{array}{c|c} \mathbb{I}_k & - \widetilde{A} \\ \hline
   - \widetilde{A}^\dagger & \mathbb{I}_k \end{array} \right) \ ,
\end{equation}
with
\begin{equation}\label{}
    \widetilde{A} = |\widetilde{\psi}_1\>\<\widetilde{\psi}_2| - |\widetilde{\psi}_2\>\<\widetilde{\psi}_1| + \<
    \widetilde{\psi}_1|\widetilde{\psi}_2\>\, \mathbb{I}_k \ ,
\end{equation}
and normalized $\widetilde{\psi}_j = \psi_j/a_j$. Hence, to show
that $\Psi^0_{2k}(P)\geq 0$ one needs to prove
\begin{equation}\label{*}
\left( \begin{array}{c|c} \mathbb{I}_k & - \widetilde{A} \\ \hline
   - \widetilde{A}^\dagger & \mathbb{I}_k \end{array} \right) \geq 0
   \ ,
\end{equation}
for arbitrary $\psi_j \neq 0$. Now, the above condition is
equivalent to
\begin{equation}\label{AAI}
\widetilde{A}\widetilde{A}^\dagger \leq \mathbb{I}_k \ .
\end{equation}
Vectors $\{\psi_1,\psi_2\}$ span 2-dimensional subspace in
$\mathbb{C}^k$ and let $\{e_1,e_2\}$ be a 2-dim. orthonormal basis
such that  $\psi_1 = e_1$ and
\begin{equation}\label{}
    \psi_2 = e^{i\lambda} s e_1 + c e_2\ ,
\end{equation}
with $s = \sin\alpha$, $c=\cos\alpha$ for some angle $\alpha$. Now,
completing the basis $\{e_1,e_2,e_3,\ldots,e_k\}$ in $\mathbb{C}^k$
one easily finds that the matrix elements of $\widetilde{A}$ has a
form of the following direct sum
\begin{equation}\label{}
    \widetilde{A} =  \left( \begin{array}{c|c} e^{-i\lambda} s & c \\ \hline
   - c & e^{i\lambda}s \end{array} \right)\, \oplus\, e^{-i\lambda}s
   \mathbb{I}_{k-2} \ .
\end{equation}
Hence
\begin{equation}\label{}
\widetilde{A}\widetilde{A}^\dagger = \mathbb{I}_2 \, \oplus \, s^2
\mathbb{I}_{k-2}\ ,
\end{equation}
which proves (\ref{AAI}) since all eigenvalues of
$\widetilde{A}\widetilde{A}^\dagger$ -- $\{1,1,s^2,\ldots,s^2\}$ --
are bounded by 1.

Now, our new positive maps can be useful in detecting entanglement
only if they are not completely positive. It is easy to check that
the corresponding Choi matrix
\begin{equation}\label{W2k}
    W_{2k} =  \sum_{i,j=1}^{2k} e_{ij} \ot \Psi^0_{2k}(e_{ij})\ ,
\end{equation}
possesses 2 negative eigenvalues $\{-1,(2-k)/k\}$ (unless $k=2$).
Hence, (\ref{W2k})  defines true entanglement witness in
$\mathbb{C}^{2k} \ot \mathbb{C}^{2k}$. As usual using Dirac notation
we define $e_{kl} := |e_k\>\<e_l|$. Note, that the corresponding
Brauer-Hall witness  possesses only one negative eigenvalue ``$-1$".
Hence these two classes are different (unless $k=2$).

\section{Properties of new entanglement witnesses}  \label{PROP}

In this section we study basis properties of $W_{2k}$.

\subsection{$W_{2k}$ are indecomposable}

To show that $W_{2k}$ is indecomposable one needs to define a PPT
state $\rho$ in  $\mathbb{C}^{2k} \ot \mathbb{C}^{2k}$ such that
${\rm Tr}(W_{2k} \rho_{2k}) < 0$. Consider the following operator
\begin{equation}\label{our-rho}
    \rho_{2k} = \sum_{i,j=1}^{2k} e_{ij} \ot \rho^{(2k)}_{ij}\ ,
\end{equation}
where the $2k \times 2k$ blocks are defined as follows: diagonal blocks
\begin{equation}\label{}
    \rho^{(2k)}_{ii} = N_k \left( \begin{array}{c|c} k\mathbb{I}_k
 & \mathbb{O}_k \\ \hline  \mathbb{O}_k & \mathbb{I}_k
\end{array} \right)\ ,
\end{equation}
for $i=1,\ldots,k$, and
\begin{equation}\label{}
    \rho^{(2k)}_{ii} = N_k \left( \begin{array}{c|c} \mathbb{I}_k
 & \mathbb{O}_k \\ \hline  \mathbb{O}_k & k\mathbb{I}_k
\end{array} \right)\ ,
\end{equation}
for $i=k+1,\ldots,2k$. The off-diagonal blocks are form:
\begin{equation}\label{}
    \rho^{(2k)}_{i,i+k} = - N_k W^{(2k)}_{i,i+k} \ ,
\end{equation}
for $i=1,\ldots,k$,
\begin{equation}\label{}
    \rho^{(2k)}_{ij} = N_k e_{ij}  \ ,
\end{equation}
for $i=1,\ldots,k$, $j=k+1,\ldots,2k$ and $j\neq i+k$ and
\begin{equation}\label{}
    \rho^{(2k)}_{ij} =  \mathbb{O}_k\ ,
\end{equation}
otherwise. The normalization factor $N_k$ is given by
\[  1/N_k = 2k^2(k+1)\ . \]
Direct calculation shows that
\[  \rho \geq 0\ , \ \ \ \rho^\Gamma \geq 0 \ , \ \ \ {\rm Tr}\rho =
1\ . \]
 For example for $k=2$ one obtains the
following  density operator
\begin{widetext}
\begin{equation}\label{}
 \rho_{4} = \frac{1}{24} \left( \begin{array}{cccc|cccc|cccc|cccc}
 2& \cdot& \cdot& \cdot& \cdot& \cdot& \cdot& \cdot& \cdot& \cdot& 1& \cdot& \cdot& \cdot& \cdot& 1\\
 \cdot& 2& \cdot& \cdot& \cdot& \cdot& \cdot& \cdot& \cdot& \cdot& \cdot& \cdot& \cdot& \cdot& \cdot& \cdot\\
 \cdot& \cdot& 1& \cdot& \cdot& \cdot& \cdot& \cdot& \cdot& \cdot& \cdot& \cdot& \cdot& \cdot& \cdot& \cdot\\
 \cdot& \cdot& \cdot& 1& \cdot& \cdot& \cdot& \cdot& \cdot& 1& \cdot& \cdot& \cdot& \cdot& \cdot& \cdot  \\ \hline
 \cdot& \cdot& \cdot& \cdot& 2 & \cdot& \cdot& \cdot& \cdot& \cdot& \cdot& \cdot& \cdot& \cdot& \cdot& \cdot \\
 \cdot& \cdot& \cdot& \cdot& \cdot& 2 & \cdot& \cdot& \cdot& \cdot& 1& \cdot& \cdot& \cdot& \cdot& 1 \\
 \cdot& \cdot& \cdot& \cdot& \cdot& \cdot& 1& \cdot& \cdot& \cdot& \cdot& \cdot& 1& \cdot& \cdot& \cdot \\
 \cdot& \cdot& \cdot& \cdot& \cdot& \cdot& \cdot& 1& \cdot& \cdot& \cdot& \cdot& \cdot& \cdot& \cdot& \cdot  \\ \hline
 \cdot& \cdot& \cdot& \cdot& \cdot& \cdot& \cdot& \cdot& 1& \cdot& \cdot& \cdot& \cdot& \cdot& \cdot& \cdot \\
 \cdot& \cdot& \cdot& 1& \cdot& \cdot& \cdot& \cdot& \cdot& 1& \cdot& \cdot& \cdot& \cdot& \cdot& \cdot \\
 1& \cdot& \cdot& \cdot& \cdot& 1& \cdot& \cdot& \cdot& \cdot& 2 & \cdot& \cdot& \cdot& \cdot& \cdot \\
 \cdot& \cdot& \cdot& \cdot& \cdot& \cdot& \cdot& \cdot& \cdot& \cdot& \cdot& 2 & \cdot& \cdot& \cdot& \cdot \\ \hline
 \cdot& \cdot& \cdot& \cdot& \cdot& \cdot& 1& \cdot& \cdot& \cdot& \cdot& \cdot& 1& \cdot& \cdot& \cdot \\
 \cdot& \cdot& \cdot& \cdot& \cdot& \cdot& \cdot& \cdot& \cdot& \cdot& \cdot& \cdot& \cdot& 1& \cdot& \cdot \\
 \cdot& \cdot& \cdot& \cdot& \cdot& \cdot& \cdot& \cdot& \cdot& \cdot& \cdot& \cdot& \cdot& \cdot& 2& \cdot \\
 1& \cdot& \cdot& \cdot& \cdot& 1& \cdot& \cdot& \cdot& \cdot& \cdot& \cdot& \cdot& \cdot& \cdot& 2 \end{array}
 \right)\ .
\end{equation}
\end{widetext}
One easily finds for the trace
\begin{equation}\label{}
    {\rm Tr}(W_{2k} \rho_{2k} ) = - \frac{k-1}{k^2(k+1)}\ ,
\end{equation}
which proves indecomposability of $W_{2k}$.

\subsection{$W_{2k}$ are atomic}

In order to prove that $W_{2k}$ is atomic one has to define a PPT
state $D_{2k}$ such that Schmidt rank of $D_{2k}$ and of its partial
transposition $D_{2k}^\Gamma$ is bounded by 2 and show that ${\rm
Tr}(W_{2k} D_{2k})< 0$. It is clear that atomicity implies
indecomposability but for clarity of presentation we treat these two
notions independently. Let us introduce the following family of
product vectors
\begin{eqnarray*}
  \phi_1 &=& e_1 \ot e_1 \ ,\\
  \phi_2 &=& e_1 \ot e_{k+1} \ , \\
  \phi_3 &=&  e_k \ot e_1 \ , \\
  \phi_4 &=& e_k \ot e_{2k} \\
  \phi_5 &=& e_{k+1} \ot e_1\ , \\
  \phi_6 &=& e_{k+1} \ot e_{k+1}\ , \\
  \phi_7 &=& e_{k+1} \ot e_{2k}\ .
\end{eqnarray*}
Define now the following positive operator
\begin{eqnarray}\label{D}
    D_{2k} &=& \frac 17 \Big( |\phi_1 + \phi_6\>\<\phi_1 + \phi_6| + |\phi_5 - \phi_4\>\<\phi_5 -
    \phi_4| \nonumber \\ &+& |\phi_2\>\<\phi_2| + |\phi_3\>\<\phi_3| +
    |\phi_7\>\<\phi_7|     \Big)\ .
\end{eqnarray}
One easily finds for its partial transposition
\begin{eqnarray}\label{DG}
    D_{2k}^\Gamma &=& \frac 17 \Big( |\phi_2 + \phi_5\>\<\phi_2 + \phi_5| + |\phi_3 - \phi_7\>\<\phi_3 -
    \phi_7| \nonumber \\ &+& |\phi_1\>\<\phi_1| + |\phi_4\>\<\phi_4| +
    |\phi_6\>\<\phi_6|     \Big)\ .
\end{eqnarray}
Now, it is clear from that both $D_{2k}$ and $D_{2k}^\Gamma$ are
constructed out of rank-1 projectors and Schmidt rank of each
projector is 1 or 2. Therefore
\[  {\rm SN}(D_{2k}) \leq 2 \ , \ \ \ {\rm SN}(D^\Gamma_{2k}) \leq 2
\ . \] Finally, one finds for the trace
\begin{equation}\label{}
    {\rm Tr}(W_{2k}D_{2k}) = - \frac{1}{7k} \ ,
\end{equation}
which shows that $W_{2k}$ defines atomic entanglement witness.

\subsection{$W_{2k}$ are optimal}

To show that $W_{2k}$ is optimal we use the following result
Lewenstein et. al. \cite{Lew1}: if the family of product vectors
$\psi \ot \phi \in \mathbb{C}^{2k} \ot \mathbb{C}^{2k}$ satisfying
\begin{equation}\label{}
    \< \psi \ot \phi| W|\psi \ot \phi \> = 0 \ ,
\end{equation}
span the total Hilbert space $\mathbb{C}^{2k} \ot \mathbb{C}^{2k}$,
then $W$ is optimal. Let us introduce the following sets of vectors:
\begin{eqnarray*}
f_{mn} = e_m + e_n \ ,
\end{eqnarray*}
and
\begin{eqnarray*}
g_{mn} = e_m + i e_n \ ,
\end{eqnarray*}
for each $1\leq m < n \leq 2k$. It is easy to check that $(2k)^2$
vectors $\psi_\alpha \ot \psi^*_\alpha$ with $\psi_\alpha$ belonging
to the set
\[
\{\, e_l\, , f_{mn}\, , g_{mn} \, \}\ , \] are linearly independent
and hence they do span $\mathbb{C}^{2k} \ot \mathbb{C}^{2k}$. Direct
calculation shows that
\begin{equation}\label{}
    \< \psi_\alpha \ot \psi^*_\alpha| W|\psi_\alpha \ot \psi^*_\alpha \> = 0 \ ,
\end{equation}
which proves that $W_{2k}$ is an optimal EW.

\subsection{$W_{2k}$ have circulant structure}

Finally, let us note that $W_{2k}$ displays so called circulant
structure \cite{PPT-nasza,CIRCULANT}. Let $\{e_1,\ldots,e_d\}$ be an
orthonormal basis in $\mathbb{C}^d$ and let $S : \mathbb{C}^d
\rightarrow \mathbb{C}^d$ be a shift operator (elementary
permutation) defined by
\begin{equation}\label{}
    S\, e_j = e_{j+1} \ , \ \ \ j=1,\ldots,d\ \ \ ({\rm mod} \ d)  \ .
\end{equation}
Now, introducing
\begin{equation}\label{}
    \Sigma_0 = {\rm span} \{ e_1\ot e_1, \ldots, e_d \ot e_d\} \ ,
\end{equation}
define
\begin{equation}\label{}
    \Sigma_\alpha = (\oper \ot S^\alpha)\Sigma_1 \ , \ \ \
    \alpha=0,\ldots,d-1\ .
\end{equation}
A bipartite operator $X : \mathbb{C}^d \ot \mathbb{C}^d \rightarrow
\mathbb{C}^d \ot \mathbb{C}^d$ displays circulant structure if
\begin{equation}\label{}
    X = X_0 \oplus \ldots \oplus X_{d-1}\ ,
\end{equation}
such that each $X_\alpha$ is supported on $\Sigma_\alpha$. It is
therefore clear that
\begin{equation}\label{}
    X_\alpha = \sum_{i,j=1}^d x^{(\alpha)}_{ij} e_{ij} \ot S^\alpha
    e_{ij} S^{\alpha\,\dagger}  \ ,
\end{equation}
where $[x^{(\alpha)}_{ij}]$ is $d \ot d$ complex matrix for each
$\alpha=0,\ldots,d-1$, i.e. a circulant bipartite operator $X$ is
uniquely defined by the collection of $d$ complex matrices
$[x^{(\alpha)}_{ij}]$.

\section{Structural physical approximation}  \label{SPA}

It is well know that positive maps cannot be directly implemented in
the laboratory. The idea of {\it structural physical approximation}
(SPA) \cite{SPA1,SPA2} is to mix a positive map $\Lambda$ with some
completely positive map making the mixture $\widetilde{\Lambda}$
completely positive. In the recent paper \cite{SPA3} the authors
analyze SPA to a positive map $\Lambda : \mathcal{\mathcal{H}_A}
\rightarrow \mathcal{\mathcal{H}_B}$ obtained through minimal
admixing of white noise
\begin{equation}\label{}
    \widetilde{\Lambda}(\rho) = p \frac{\mathbb{I}_B}{d_B} \, {\rm
    Tr}(\rho) + (1-p) \Lambda(\rho)\ .
\end{equation}
The minimal means that the positive mixing parameter $0 < p < 1$ is
the smallest one for which the resulting map $\widetilde{\Lambda}$
is completely positive, i.e. it defines a quantum channel.
Equivalently, one may introduce SPA of an entanglement witness $W$:
\begin{equation}\label{}
    \widetilde{W} = \frac{p}{d_A d_B} \mathbb{I}_A \ot \mathbb{I}_B
    + (1-p) W \ ,
\end{equation}
where $p$ is the smallest parameter for which $\widetilde{W}$ is a
positive operator in $\mathcal{H}_A \ot \mathcal{H}_B$, i.e. it
defines (possibly unnormalized) state.

It was conjectured \cite{SPA3} that SPA to optimal positive maps
correspond to entanglement breaking maps (channels). Equivalently,
SPA to optimal entanglement witnesses correspond to separable
(unnormalized) states. It turns out that the family of optimal
maps/witnesses constructed in this paper does support this
conjecture.

The corresponding SPA of $W_{2k}$ is given by
\begin{equation}\label{}
    \widetilde{W}_{2k} = \frac{p}{(2k)^2} \mathbb{I}_{2k} \ot
    \mathbb{I}_{2k}     + (1-p) W_{2k} \ .
\end{equation}
Using the fact that the maximal negative eigenvalue of $W_{2k}$
equals ``$-1$'' one easily finds the following condition for the
positivity of $\widetilde{W}_{2k}$
\begin{equation}\label{p}
    p \geq \frac{d}{d+1}\ ,
\end{equation}
with $d = 2k$. Surprisingly, one obtains the same bound for $p$ as
in Eqs. (26), (33) and (65) in \cite{SPA3}.

Now, to show that SPA of $\Psi^0_{2k}$ (or eqivalently $W_{2k}$) is
entanglement breaking (equivalently separable) we use the following

\begin{lemma} \cite{AA} Let $\Lambda : \mathcal{B}(\mathbb{C}^d) \rightarrow
\mathcal{B}(\mathbb{C}^d)$ be a positive unital map. Then SPA of
$\Lambda$ is entanglement breaking if $\Lambda$ detects all
entangled isotropic states in $\mathbb{C}^d \ot \mathbb{C}^d$.
\end{lemma}
Indeed, let
\begin{equation}\label{}
    \rho_p = \frac{p}{d^2}\, \mathbb{I}_d \ot \mathbb{I}_d + (1-p)
    P^+_d\ ,
\end{equation}
be an isotropic state which is known to be entangled iff
\begin{equation}\label{p-iso}
p < \frac{1}{d+1}\ .
\end{equation}
 Now, assume that $(\oper \ot \Lambda) \rho_p$ is not
positive if $\rho_p$ is entangled. Using $\Lambda(\mathbb{I}_d) =
\mathbb{I}_d$ one obtains
\begin{equation}\label{}
(\oper \ot \Lambda) \rho_p = \frac{p}{d^2}\, \mathbb{I}_d \ot
\mathbb{I}_d + (1-p) W\ ,
\end{equation}
that is, $(\oper \ot \Lambda) \rho_p  = \widetilde{W}$. Now, if
$\widetilde{W}$ is positive, then $\rho_p$ has to be separable
(otherwise it would be detected by $\Lambda$). But since $\oper \ot
\Lambda$ sends separable states into separable states one concludes
that $\widetilde{W}$ is separable (or equivalently
$\widetilde{\Lambda}$ is entanglement breaking).

\begin{lemma}
If in addition $\Lambda$ is self-dual, i.e.
\begin{equation}\label{self}
    {\rm Tr}(A \cdot \Lambda(B)) = {\rm Tr}(\Lambda(A) \cdot B)\ ,
\end{equation}
for all $A,B \in \mathcal{B}(\mathbb{C}^d)$, then it is enough to
check whether all entangled isotropic states are detected by the
witness $W$, i.e. ${\rm Tr}(W \rho_p) < 0$ for all $p$ satisfying
(\ref{p-iso}).
\end{lemma}
Again, the proof is very easy. One has
\begin{eqnarray}\label{}
    {\rm Tr}(\rho_p \cdot W) &=&  {\rm Tr}(\rho_p \cdot (\oper \ot
    \Lambda)P^+_d)\nonumber \\  &=& {\rm Tr}((\oper \ot
    \Lambda)\rho_p \cdot P^+_d)\ ,
\end{eqnarray}
where in the last equality we used the self-duality of $\Lambda$.
Now, if ${\rm Tr}(\rho_p \cdot W) < 0$ for $p$ satisfying
(\ref{p-iso}), then $(\oper \ot \Lambda)\rho_p$ is not positive
(otherwise its trace with the projector $P^+_d$ would be positive).
Hence by Lemma 1 SPA $\widetilde{\Lambda}$ is entanglement breaking.

We are prepared to show that SPA for $\Psi^0_{2k}$ is entanglement
breaking.
\begin{lemma} $\Psi^0_{2k}$ is self-dual.
\end{lemma}
One checks by direct calculations that
\begin{equation}\label{}
    {\rm Tr}(e_{kl} \cdot \Psi^0_{2k}(e_{mn})) = {\rm Tr}(\Psi^0_{2k}(e_{kl}) \cdot e_{mn})\ ,
\end{equation}
for all $k,l,m,n=1,\ldots,2k$. Hence, due to the Lemma 2, to show
that SPA for $\Psi^0_{2k}$ is entanglement breaking is it enough to
prove
\begin{lemma} ${\rm Tr}(W_{2k} \rho_p) < 0$ for all $p$ satisfying
(\ref{p-iso}).
\end{lemma}
To prove it let us note that
\begin{eqnarray*}\label{}
    {\rm Tr}(W_{2k} \rho_p) = \frac{p}{(2k)^2} {\rm Tr}\,W_{2k} + (1-p)
    {\rm Tr}(W_{2k} P^+_{2k})\ .
\end{eqnarray*}
Now, ${\rm Tr}\,W_{2k}=2k$, and
\begin{equation*}\label{}
{\rm Tr}(W_{2k} P^+_{2k}) = \frac{1}{2k} \sum_{m,n=1}^{2k} \<
m|\Psi^0_{2k}(e_{mn})|n\> \ .
\end{equation*}
Finally, using definition of $\Psi^0_{2k}$ one gets
\begin{equation*}\label{}
\sum_{m,n=1}^{2k} \< m|\Psi^0_{2k}(e_{mn})|n\>  = - 2k\ ,
\end{equation*}
and hence
\begin{equation}\label{}
{\rm Tr}(W_{2k} \rho_p) = \frac{p(d+1) -1}{d}\ ,
\end{equation}
with $d=2k$. Now, if $\rho_p$ is entangled, i.e. $p < 1/(d+1)$, then
${\rm Tr}(W_{2k} \rho_p) < 0$ which shows that $W_{2k}$ detects all
entangled isotropic states.

\section{Conclusions}

We have provided a new construction of EWs in $\mathbb{C}^d \ot
\mathbb{C}^d$ with $d=2k$. It was shown that these EWs are
indecomposable, i.e. they are able to detect PPT entangled state.
Moreover, they are so called atomic EWs, i.e. they able to detect
PPT entangled states $\rho$ such that both $\rho$ and $\rho^\Gamma$
possess Schmidt number 2. The crucial property of our witnesses is
they optimality, i.e. they are no other witnesses which can detect
more entangled states.

Equivalently, our construction gives rise to the new class of
positive maps in algebras of $d \times d$ complex matrices. For
$k=2$ this construction reproduces old example of Robertson map
\cite{Robertson} and hence \cite{atomic} defines the special case of
Brauer-Hall maps \cite{Breuer,Hall}.

Let us observe that if $\Lambda : \mathcal{B}(\mathbb{C}^d)
\rightarrow \mathcal{B}(\mathbb{C}^d)$ is a positive indecomposable
map then for any unitaries $U_1,U_2 : \mathbb{C}^d \rightarrow
\mathbb{C}^d$ a new map
\begin{equation}\label{}
    \Lambda^{U_1U_2}(A) := U_1 \Lambda(U_2^\dagger A
    U_2)U_1^\dagger\ ,
\end{equation}
is again positive and indecomposable \cite{atomic}. This observation
enables one to generalize Robertson map $\Phi^0_{2k}$ and our new
map $\Psi^0_{2k}$ to $\Phi^{U_1U_2}_{2k}$ and $\Psi^{U_1U_2}_{2k}$.
Note, that if $U_1 = U_2 = U$ given by (\ref{BH}), then
$\Phi^{U}_{2k} := \Phi^{UU}_{2k}$ defines Breuer-Hall map.
Therefore, $\Psi^{U}_{2k} := \Psi^{UU}_{2k}$ may be regarded as a
Breuer-Hall-like generalization of our primary map $\Psi^0_{2k}$.

It should be stressed , an EW defined by the Breuer-Hall map and EW
$W_{2k}$ introduces in this paper are different, i.e. they do detect
different classes of PPT entangled states. Direct calculation shows
that the PPT entangled state (\ref{our-rho}) is not detected by the
Breuer-Hall witness. On the other hand consider the family of PPT
entangled state introduced in \cite{Breuer}
\begin{equation}\label{lambda}
    \rho(\lambda) = \lambda P^+_d + (1-\lambda) \rho_0\ ,
\end{equation}
with
\begin{equation}\label{}
    \rho_0 = \frac{2}{d(d+1)}\, U_0 P_S U_0^\dagger\ ,
\end{equation}
and $P_S$ being the projector onto the subspace of states symmetric
under the swap operation. It was shown that $\rho(\lambda)$ is PPT
for $0 \leq \lambda \leq 1(d+2)$. Moreover, Breuer-Hall witness
detects all entangled states within $\lambda$-family (\ref{lambda})
(both PPT and NPT). Direct calculation shows that our witness
$W_{2k}$ does not detect PPT entangled states in (\ref{lambda}).

Interestingly, the partial transposition $W_{2k}^\Gamma$ defines an
EW with $k(k-1)/2$ negative eigenvalues (all equal to `$-1$'). For
$k=2$ it gives exactly one negative eigenvalue (the fact well known
from the family of Breuer-Hall maps in 4 dimensions). Therefore,
this example provide an EW with multiple negative eigenvalues.
However, contrary to the Breuer-Hall maps we were not able to show
that $W_{2k}^\Gamma$ is optimal.

We have shown that structural physical approximation for our new
class of positive maps gives rise to entanglement breaking channels.
This result  supports recent conjecture by Korbicz et. al.
\cite{SPA3}.

Finally, let us mention some open questions. In this paper we have
used reduction map as a building block to construct other optimal
maps. Can we use other positive maps as building blocks? Is it true
that properties of  building blocks (like optimality and/or
atomicity) are shared by the map which is built out of them?

\section*{Acknowledgement}
We thank Antonio Acin and Andrzej Kossakowski for valuable
discussions. This work was partially supported by the Polish
Ministry of Science and Higher Education Grant No 3004/B/H03/2007/33
and by the Polish Research Network  {\it Laboratory of Physical
Foundations of Information Processing}.


\begin{thebibliography}{1} \bibliographystyle{plain}

\bibitem{QIT} M. A. Nielsen and I. L. Chuang, Quantum
Computation and Quantum Information, (Cambridge University Press,
Cambridge, England, 2000).

\bibitem{HHHH} R. Horodecki, P. Horodecki, M. Horodecki and K.
Horodecki, Rev. Mod. Phys. {\bf 81}, 865 (2009).

\bibitem{EW1}  M. Horodecki, P. Horodecki and R. Horodecki, Phys. Lett. A {\bf 223}, 1
(1996).

\bibitem{EW2} B.M. Terhal, Phys. Lett. A
{\bf 271}, 319 (2000).


\bibitem{Terhal2} B. M. Terhal, Theor. Comput. Sci. {\bf 287}, 313 (2002).


\bibitem{O} M. Lewenstein, B. Kraus, J. I. Cirac, and P. Horodecki, Phys. Rev. A {\bf 62}, 052310
(2000).

\bibitem{Lew1} M. Lewenstein, B. Kraus, P. Horodecki, and J. I. Cirac, Phys. Rev.
A {\bf 63}, 044304 (2001).


\bibitem{Lew2} B. Kraus, M. Lewenstein, and J. I. Cirac, Phys. Rev. A {\bf 65}, 042327
(2002).


\bibitem{Lew3} P. Hyllus, O. G\"uhne, D. Bruss, and M. Lewenstein
Phys. Rev. A {\bf 72}, 012321 (2005).

\bibitem{Bruss} D. Bru\ss, J. Math. Phys. {\bf 43}, 4237 (2002).

\bibitem{Toth} G. T\'oth and O. G\"uhne, Phys. Rev. Lett. {\bf 94}, 060501 (2005)

\bibitem{Bertlmann} R. A. Bertlmann, H. Narnhofer and W. Thirring,
Phys. Rev. A {\bf 66}, 032319 (2002).

\bibitem{Brandao} F.G.S.L. Brand{\~a}o, Phys. Rev. A {\bf 72}, 022310 (2005).

\bibitem{Gniewko} G. Sarbicki, J. Phys. A:
Math. Theor.  {\bf 41},  375303 (2008).

\bibitem{how} D. Chru\'sci\'nski and A. Kossakowski, J. Phys. A:
Math. Theor.  {\bf 41}, 145301  (2008).

\bibitem{EX} M. Bourennane, M. Eibl, C. Kurtsiefer, S. Gaertner, H.
Weinfurter, O. G\"uhne, P. Hyllus, D. Bruss, M. Lewenstein and A.
Sanpera, Phys. Rev. Lett. {\bf 92}, 087902 (2004).

\bibitem{Wu} L.-A. Wu, S. Bandyopadhyay, M. S. Sarandy, and D. A. Lidar, Phys. Rev. A {\bf 72}, 032309
(2005).

\bibitem{O1} A.C. Doherty, P.A. Parrilo and F.M. Spedalieri, Phys.
Rev. Lett. {\bf 88}, 187904 (2002).

\bibitem{O2} F.G.S.L. Brand{\~a}o and R.O. Vianna, Phys. Rev. Lett.
{\bf 93}, 220503 (2004).

\bibitem{O3} J. Eisert, P. Hyllus, O. G\"uhne, and M. Curty, Phys. Rev.
A {\bf 70}, 062317 (2004);

\bibitem{W-Bell} P. Hyllus, O. Guehne, D. Bru\ss, M. Lewenstein, Phys. Rev. A {\bf
72}, 012321 (2005).



\bibitem{Choi} M.-D. Choi,  Lin. Alg. Appl. {\bf 10}, 285 (1975).

\bibitem{Jam} A. Jamio{\l}kowski, Rep. Math. Phys. {\bf 3}, 275 (1972).


\bibitem{Woronowicz1} S.L. Woronowicz, Rep. Math. Phys. {\bf 10}, 165
(1976).






\bibitem{Stormer1}  E. St{\o}rmer, Acta Math. {\bf 110}, 233
(1963).

\bibitem{Choi1} M.-D. Choi,  Lin. Alg. Appl. {\bf 12}, 95 (1975);  M.-D. Choi, J. Operator Theory, {\bf 4}, 271
(1980).

\bibitem{Woronowicz2} S.L. Woronowicz, Comm. Math. Phys. {\bf 51}, 243 (1976).

\bibitem{Robertson} A.G. Robertson,   Math. Proc. Camb.  Phil.  Soc., {\bf 94}, 71 (1983).
{\bf 34}, 87 (1983); A.G. Robertson, J. London Math. Soc. (2) {\bf
32}, 133 (1985).

\bibitem{Tang} W.-S. Tang, Lin. Alg. Appl. {\bf 79}, 33 (1986).

\bibitem{TT} K. Tanahashi and J. Tomiyama, Canad. Math. Bull.
{\bf 31}, 308 (1988).

\bibitem{Osaka} H. Osaka,  Lin. Alg. Appl. {\bf 153}, 73 (1991); {\em ibid}. {\bf 186}, 45
(1993).


\bibitem{Benatti} F. Benatti, R. Floreanini and M. Piani, Phys.
Lett. A {\bf 326}, 187 (2004).

\bibitem{Ha} K.-C. Ha, Publ. RIMS, Kyoto Univ. {\bf 34}, 591
(1998)


\bibitem{Kye} K.-C. Ha and S.-H. Kye,  J. Phys. A: Math. Gen. {\bf 38}, 9039
(2005); Phys. Lett. A {\bf 325}, 315 (2004).

\bibitem{Kossak1} A. Kossakowski, Open Sys. Information Dyn. {\bf 10},
213 (2003).

\bibitem{Breuer} H.-P. Breuer, Phys. Rev. Lett. {\bf 97}, 0805001
(2006).

\bibitem{Hall} W. Hall, J. Phys. A: Math. Gen. {\bf 39}, (2006)
14119.



\bibitem{OSID-W} D. Chru\'sci\'nski and A. Kossakowski,  Open Systems and Inf.
Dynamics, {\bf 14}, 275 (2007).

\bibitem{atomic} D. Chru\'sci\'nski and A. Kossakowski, J. Phys. A:
Math. Theor.  {\bf 41}, 215201  (2008).

\bibitem{CMP} D. Chru\'sci\'nski and A. Kossakowski, {\em Spectral conditions for positive
maps}, arXiv:0809.4909  (to be published in Comm. Math. Phys.)

\bibitem{kule} D. Chru\'sci\'nski and A. Kossakowski, Phys. Lett. A
{\bf 373}, 2301 (2009).


\bibitem{SPA1} P. Horodecki, Phys. Rev. A {\bf 68}, 052101 (2003).


\bibitem{SPA2} P. Horodecki and A. Ekkert, Phys. Rev. Lett. {\bf
89}, 127902 (2002).

\bibitem{SPA3} J.K. Korbicz, M.L. Almeida, J. Bae, M. Lewenstein and
A. Acin, Phys. Rev. A {\bf 78}, 062105 (2008).







\bibitem{SN} B. Terhal and P. Horodecki,
Phys. Rev. A {\bf 61}, 040301 (2000);  A. Sanpera, D. Bru{ss} and M.
Lewenstein,  Phys. Rev. A {\bf 63}, 050301(R) (2001)


\bibitem{Horodecki} M. Horodecki and P. Horodecki, Phys. Rev. {\bf A} 59, 4206
(1999).

\bibitem{PPT-nasza} D. Chru\'sci\'nski and A Kossakowski, Phys.
Rev. A {\bf 74}, 022308 (2006)

\bibitem{CIRCULANT} D. Chru\'sci\'nski and A Kossakowski, Phys.
Rev. A {\bf 76}, 032308 (2007); D. Chru\'sci\'nski and A. Pittenger,
J. Phys. A: Math. Theor. {\bf 41},  385301 (2008).


\bibitem{AA} We thank Antonio Acin for his remarks.


\end{thebibliography}
\end{document}